\documentclass[%
 preprint,
 amsmath,amssymb,
 aps,
pre
]{revtex4-1}

\usepackage{amsmath}
\usepackage{amssymb}

\usepackage{url}
\usepackage{CJKutf8}

\usepackage{array}
\newcolumntype{P}[1]{>{\raggedleft\arraybackslash}p{#1}}
\usepackage{booktabs}

\usepackage{type1cm}
\usepackage{graphicx}
\usepackage{epstopdf}
\usepackage{dcolumn}
\usepackage{bm}
\usepackage[usenames,dvipsnames]{xcolor} 

\newcounter{figcountSI}

\newcommand{\figcountSI}{\refstepcounter{figcountSI}}

\newcounter{tabcountSI}

\newcommand{\tabcountSI}{\refstepcounter{tabcountSI}}

\newcounter{seccountSI}

\newcommand{\seccountSI}{\refstepcounter{seccountSI}}

\begin{document}
\title{Fragmenting networks by targeting collective influencers at a mesoscopic level}

\author{Teruyoshi Kobayashi}
 \affiliation{Graduate School of Economics, Kobe University, Kobe, Japan}
\author{Naoki Masuda}%
 \email{Corresponding author: naoki.masuda@bristol.ac.uk}
\affiliation{Department of Engineering Mathematics, University of Bristol, Bristol, UK}%



\begin{abstract}
A practical approach to protecting networks against epidemic processes such as spreading of infectious diseases, malware, and harmful viral information is to remove some influential nodes beforehand to fragment the network into small components. Because determining the optimal order to remove nodes is a computationally hard problem, various approximate algorithms have been proposed to efficiently fragment networks by sequential node removal. Morone and Makse proposed an algorithm employing the non-backtracking matrix of given networks, which outperforms various existing algorithms. In fact, many empirical networks have community structure, compromising the assumption of local tree-like structure on which the original algorithm is based. We develop an immunization algorithm by synergistically combining the Morone-Makse algorithm and coarse graining of the network in which we regard a community as a supernode. In this way, we aim to identify nodes that connect different communities at a reasonable computational cost. The proposed algorithm works more efficiently than the Morone-Makse and other algorithms on networks with community structure. \\
%
{\bf{keywords}}: network; community structure; epidemics
\end{abstract}
\maketitle

\section*{Introduction}

Identification of influential nodes in a network is a 
topic of interest in network analysis, enjoying numerous applications.
For example, a removal or immunization of an influential node may suppress spreading of an infectious disease that may occur later. A viral information spreading campaign starting from an influential node may be more successful than a campaign starting from other nodes. There are various notions of influential nodes, as evinced by a multitude of definitions of node's centrality corresponding to the aforementioned and other applications \cite{Newman2010book}. Among them, a major criterion of the influential node is that the removal of a node, or immunization, efficiently fragments the network into small pieces. Because the problem of finding the minimal set of nodes to be immunized to fragment the network is NP-hard \cite{Altarelli2014PRX}, various immunization algorithms to determine the order of the nodes to be removed to realize efficient fragmentation of the network have been proposed~\cite{Restrepo2008,ChenPaul2008PRL,Masuda2009NJP,Salathe2010PlosCB,Schneider2011PRE,Altarelli2014PRX,Zhao2014PLOS,Morone2015Nature,Zahedi2015IKT,Requiao2015Plosone,Mugisha2016PRE}, sometimes with the constraint that the information about the network is only partially available~\cite{Cohen2003PhysRevLett-immu,Gallos2007PRER,Salathe2010PlosCB, Gong2013Plosone,Hebert2013Scirep,Liu2016Chaos}. Notably, although immunizing hubs (i.e., nodes with a large degree) first is intuitive and much better than randomly selecting nodes to be immunized \cite{Albert2000Nature,Callaway2000PhysRevLett,Cohen2001PhysRevLett}, many immunization algorithms outperform the hub-first immunization algorithm.

Morone and Makse proposed a scalable and powerful algorithm to sequentially remove nodes and fragment the network into small components as early as possible~\cite{Morone2015Nature}. Founded on the message passing approach and theory of non-backtracking matrices, the method calculates the so-called collective influence (CI) for each node to rank the nodes for prioritization. 
Their method, which is referred to as the CI algorithm, outperforms various other known methods in model and empirical networks. In the present study, we propose a new CI-based immunization algorithm that is designed to perform well when the network has community structure.

The CI algorithm assumes that the given network is locally tree-like.
In fact, a majority of empirical networks are not locally tree-like.
At a microscopic level, empirical networks are usually clustered, i.e., full of triangles~ \cite{Newman2010book}. At a mesoscopic level, many networks are composed of communities such that links are dense within communities and sparse across different communities \cite{Fortunato2010PhysRep}. Although the CI algorithm also seems to work efficiently in loopy networks unless loops are not excessive \cite{Morone2015Nature}, the performance of the CI algorithm on networks with community structure is unclear.
Some extant immunization algorithms are explicitly or implicitly informed by community structure~\cite{Holme2002PRE_attack,Ueno2008JTB,ChenPaul2008PRL,Masuda2009NJP,Salathe2010PlosCB,Gong2013Plosone,Hebert2013Scirep,Zahedi2015IKT}.
The immunization algorithms using the betweenness centrality are effective on networks with community structure~\cite{Holme2002PRE_attack,Ueno2008JTB,Salathe2010PlosCB,Schneider2011PRE,Hebert2013Scirep}. However, they are not scalable due to a high computational cost of calculating the betweenness centrality
\cite{Brandes2001JMathSociol}.
For other immunization algorithms exploiting community structure of networks, their performance relative to the CI algorithm is unknown in general \cite{Masuda2009NJP,Zahedi2015IKT} or at least for networks with community structure \cite{ChenPaul2008PRL,Morone2015Nature}.
Yet other community-based immunization algorithms impose that only local information about the network is available, mimicking realistic constraints \cite{Salathe2010PlosCB,Gong2013Plosone,Hebert2013Scirep}.
This constraint naturally limits the performance of an immunization algorithm.

We develop an immunization algorithm by formulating a CI algorithm for a coarse-grained network, in which a node represents a community, and a weighted link represents the number of links between different communities. 
We compare the performance of the proposed algorithm with that of the CI algorithm \cite{Morone2015Nature}, and the conventional algorithm targeting hubs \cite{Albert2000Nature,Callaway2000PhysRevLett,Cohen2001PhysRevLett}, and
others~\cite{Masuda2009NJP,Zahedi2015IKT} when networks have community structure.

\section*{Theory}

Consider an undirected and unweighted network having $N$ nodes.
The aim of an immunization algorithm is to sequentially remove nodes to fragment  the network as soon as possible, i.e., with a small number of removed nodes.

\subsection*{Collective influence\label{sub:CI}}
 
The CI algorithm is based on the scoring of nodes according to the CI value~\cite{Morone2015Nature}. The CI of node $i$ is defined as
\begin{align}\label{eq:CI}
  \text{CI}_\ell(i) = z_i\sum_{j\in \partial \text{Ball}(i,\ell)}z_j, 
\end{align}
where
\begin{equation}
z_i\equiv k_i-1,
\end{equation}
$k_i$ is the degree of node $i$, and $\partial\text{Ball}(i,\ell)$ is the set of nodes at distance $\ell$ from node $i$. When $\ell=0$, the CI is equivalent to the degree as long as the rank order is concerned.

The CI algorithm calculates the $\text{CI}_{\ell}(i)$ value of all nodes and removes the node with the largest CI value
%
%
in one step. Then, the CI values of all the remaining nodes are recalculated, and the same procedure is repeated. 
%
 
In fact, we use the order of nodes to be removed determined above as a tentative order. To improve the overall performance, we reorder the nodes by reinserting them as follows.
We start from the situation in which the fraction
of nodes in the largest connected component (LCC) is equal to or less than 0.01 for the first time. Then, we calculate for each removed node $i$ the number of components that node $i$ connects if it is reinserted in the current network. Next, we add back the node that connects the smallest number of connected components. We repeat this procedure until all the removed nodes are reinserted such that the initial network is restored.

The computation time of the CI algorithm is evaluated as follows \cite{Morone2015Nature}.
The calculation of $\text{CI}_\ell(i)$ requires $O(1)$ time for one node, and hence $O(N)$ time for all nodes. Because sorting the $\text{CI}_\ell(i)$ values consumes $O(N\log N)$ time, each step of the CI algorithm consumes $O(N\log N)$ time.
Therefore, the total computation time until $O(N)$ nodes are removed is evaluated as $O(N^2 \log N)$.
%
%
However, by exploiting the fact that the CI values of only $O(1)$ nodes are affected by the removal of a single node, one can accelerate the same algorithm with a max-heap data structure, yielding $O(N\log N)$ total computation time~\cite{Morone2016SciRep}.  

\subsection*{Community-based collective influence\label{sub:CbCI}}

Community structure may make a network not locally tree-like. We propose an immunization algorithm by running a weighted-network variant of the CI algorithm on a coarse-grained network in which a community constitutes a supernode. We first run a community detection algorithm. Denote by $N_{\rm{C}}$ the number of communities and by $\tilde{A}$ the $N_{\rm{C}}\times N_{\rm{C}}$ coarse-grained weighted adjacency matrix whose $(I,J)$ element is equal to the number of links that connect communities $I$ and $J$ ($I \neq J$). We use lowercases (e.g., $i$, $j$) to denote individual nodes and uppercases (e.g., $I$, $J$) to denote supernodes, i.e., communities, throughout the text. The diagonal elements of $\tilde{A}$ are set to zero.
  
Assume that the coarse-grained network is locally tree-like. By taking into account the fact that the coarse-grained network is generally a weighted network, we define the CI of community $I$ in the coarse-grained network by 
\begin{align}\label{eq:communityCIpre}
 \text{CI}_{\ell}^{\prime}(I) = z_I^{\prime} \sum_{J\in \tilde{\partial} \text{Ball}(I,{\ell})}z_{J}^{\prime\prime},
\end{align}
where $\tilde{\partial}$Ball$(I,\ell)$ denotes the set of the communities whose distance from community $I$ is equal to $\ell$ in the coarse-grained network.

We set
\begin{equation}
z_I^{\prime} \equiv \sum_{I^{\prime}=1}^{N_{\rm C}} \tilde{A}_{II^{\prime}} - \min_{I^{\prime}}\tilde{A}_{II^{\prime}}.
\label{eq:tildez}
\end{equation}
This definition is analogous to $z_i\equiv k_{i}-1$ in Eq.~\eqref{eq:CI}.
With this definition of $z_{I}^{\prime}$, the CI of community $I$ is equal to zero when $I$ has only one neighbor, as in the original CI~\cite{Karrer2014,Morone2015Nature}.

\begin{figure}
\includegraphics[width=1\columnwidth,clip]{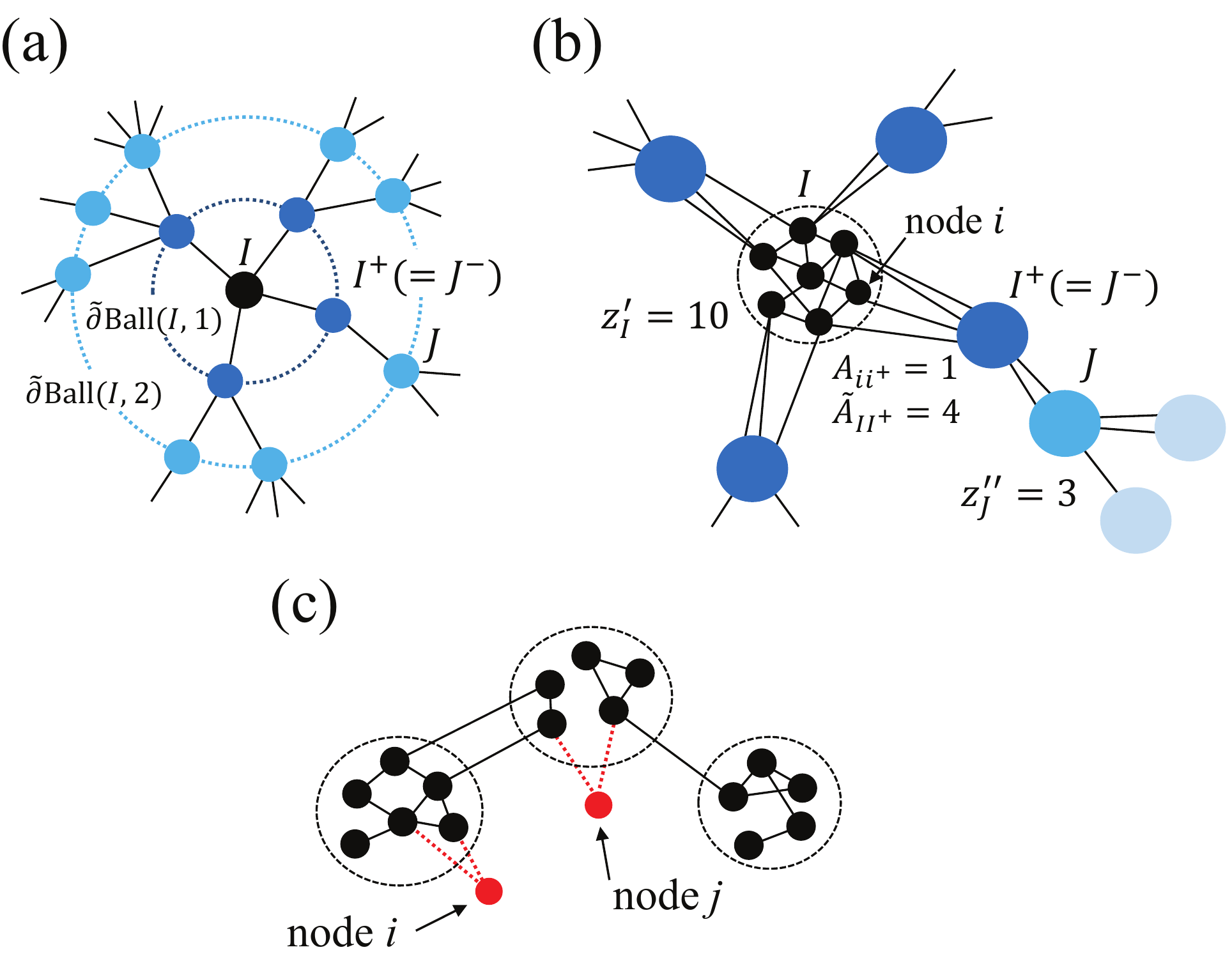}
\caption{Concept of the community-based collective influence. (a) Egocentric view of the coarse-grained network. Each circle represents a community. Two communities are adjacent by a weighted link if a node in one community is connected to at least one node in the other community. The link weight in the coarse-grained network is equal to the number of links that connect the two communities in the original network. Local tree-like structure of the coarse-grained network is assumed. (b) Illustration of $z_I^{\prime}$ and $z_{J}^{\prime\prime}$ for $\ell = 2$, in which case $I^+ = J^-$. A line represents a link in the original network. The dashed circle represents the $I$th community. (c) Schematic of community-based reinsertion. A dashed circle represents a community. Suppose that we will reinsert either node $i$ or $j$. If reinserted, node $i$ and $j$ would have a path to two and three communities, respectively. Therefore, we reinsert node $i$.}
\label{fig:schematic}
\end{figure}

We set
\begin{equation}
z_{J}^{\prime\prime} \equiv \sum_{J^{\prime}=1}^{N_{\rm C}}\tilde{A}_{JJ^{\prime}} - \tilde{A}_{J J^-}\quad (\ell\ge 1),
\label{eq:hatz}
\end{equation}
where $J$ is a community that is at distance $\ell$ from $I$, and $J^-$ is the community that is at distance ${\ell}-1$ from $I$ and on the path between $I$ and $J$ (Fig.~\ref{fig:schematic}(a)). 
It should be noted that $z_{J}^{\prime\prime}$ is equal to zero if $J^-$ is the only neighbor of $J$.
It should also be noted that, when every community consists of only one node in the original network, ${\text{CI}_{\ell}^{\prime}}(i) = \text{CI}_{\ell}(i)$ for $1\le i\le N$. 
Equation~\eqref{eq:hatz} is ill-defined for $\ell=0$. To be consistent with the original definition of the CI, we define $z_J^{\prime\prime}\equiv z_J^{\prime}$ for $\ell=0$. Then, $\text{CI}_0^{\prime}(I)$ is large when node $I$ has a large degree in the coarse-grained network.

Let $A=(A_{ij})$ be the adjacency matrix of the original network.
Equation~\eqref{eq:communityCIpre} is rewritten as 
\begin{align}\label{eq:communityCI}
 {\text{CI}_{\ell}^\prime}(I) = z_I^{\prime}\sum_{i\in \text{ community } I}
 \sum_{I^{\prime}\in \tilde{\partial}\text{Ball}(I,1)}
\frac{\sum_{i^{\prime}\in I^{\prime}}{A}_{ii^{\prime}}}{\tilde{A}_{II^{\prime}}}
 \sum_{\begin{subarray}{c}J\in \tilde{\partial} \text{Ball}(I,{\ell})\\
 I^+ = I^{\prime}\end{subarray}}
 z_{J}^{\prime\prime},
\end{align}
where $I^+$ is the community adjacent to $I$ (hence distance one from $I$) through which $J$ is reached from $I$ (Fig.~\ref{fig:schematic}(b)).
On the basis of Eq.~\eqref{eq:communityCI}, we define the community-based collective influence (CbCI) of node $i$, denoted by $\text{CbCI}(i)$, as
\begin{align}\label{eq:nodeCbCI}
 {\text{CbCI}}(i) = 
 z_I^{\prime} \sum_{I^{\prime}\in \tilde{\partial}\text{Ball}(I,1)}
\frac{\sum_{i^{\prime}\in I^{\prime}}{A}_{ii^{\prime}}}{\tilde{A}_{II^{\prime}}}
 \sum_{\begin{subarray}{c}J\in \tilde{\partial} \text{Ball}(I,{\ell})\\
 I^+ = I^{\prime}\end{subarray}}
 z_{J}^{\prime\prime},
\end{align}
where node $i$ belongs to community $I$.
In Eq.~\eqref{eq:nodeCbCI}, the importance of a node stems from three factors. First, ${\text{CbCI}}(i)$ is proportional to $z_I^{\prime}$, which is essentially the number of inter-community links of the community to which $i$ belongs. Second, 
${\text{CbCI}}(i)$ is large if $I$ has many high-degree nodes at distance ${\ell}$ in the coarse-grained network (i.e., sum of $z_{J}^{\prime\prime}$). Third, ${\text{CbCI}}(i)$ is large if node $i$ has many inter-community links relative to the total number of inter-community links that community $I$ has
(i.e., $\sum_{i^{\prime}\in I^{\prime}}{A}_{ii^{\prime}}/\tilde{A}_{II^{\prime}}$).
We set ${\ell}=2$ in the following numerical simulations.
When $\ell=2$, $I^+$ in Eqs.~\eqref{eq:communityCI} and \eqref{eq:nodeCbCI} coincide with $J^-$ in Eq.~\eqref{eq:hatz} (Fig.~\ref{fig:schematic}(b)).

We remove the node with the largest CbCI value.
If there are multiple nodes with the same largest CbCI value, we select the node having the largest degree. If there are multiple nodes with the same largest CbCI and degree, we break the tie at random. Then, we recalculate the CbCI for all remaining nodes, remove the node with the largest CbCI, and repeat the same procedure until the size of the LCC becomes equal to or less than $0.01N$. We further optimize the obtained order of node removal by reinsertion, as in the CI algorithm. We use the coarse-grained network, not the original network, to inform the reinsertion process in the CbCI algorithm. In other words, the number of communities that belong to the same component as the reinserted node is measured for each tentatively reinserted node. We decide to reinsert the node whose presence connects the least number of communities (Fig.~\ref{fig:schematic}(c)).

Given a partitioning of the network into communities, the calculation of $\text{CbCI}(i)$ for one node consumes $O(1)$ time. Therefore, if we adapt the original implementation of the CI algorithm \cite{Morone2015Nature} to the case of the CbCI, sorting of $\text{CbCI}(i)$ dominates the computation time of the CbCI algorithm. The time complexity of the CbCI algorithm is the same as that of the CI algorithm in Ref.~\cite{Morone2015Nature}, i.e., $O(N^2\log N)$, if community detection is not a bottleneck.
The use of the max-heap data structure makes the CbCI algorithm run in $O(N\log N)$ time if $N_{\rm C} = O(N)$ such that the CbCI values of $O(1)$ nodes are affected by the removal of a single node. Generally speaking, the CbCI algorithm with the max-heap data structure runs in $O(N\log N)\times O(N/N_{\rm C}) = O((N^2/N_{\rm C})\log N)$ time.

We use the following six algorithms for community detection: (i) Infomap \cite{Rosvall2008, Rosvall2010}, requiring
$O(M)$ time \cite{Fortunato2010PhysRep}, where $M$ is the number of links, and hence $O(N)$ time for sparse networks; (ii) Walktrap, which requires $O(N^2\log N)$ for most empirical networks~\cite{Pons2005}; (iii) the label-propagation algorithm, requiring nearly linear time in $N$~\cite{Raghavan2007};
(iv) a fast greedy algorithm for modularity maximization, requiring $O(N(\log N)^2)$ time for sparse networks~\cite{Clauset2004}; (v) modularity maximization based on simulated annealing, which is practical up to $\approx 10^4$ nodes in the original paper~\cite{Sales-Pardo2007} and time-consuming because modularity must be maximized in a parameter-dependent manner \cite{Lancichinetti2009PRE};
%
%
(vi) the Louvain algorithm, which practically runs in $O(N)$ time~\cite{Blondel2008}. %
%
The last three algorithms intend to maximize the modularity, denoted by $Q$. The first three algorithms detect communities according to different criteria.

Except for the simulated annealing algorithm, the computational cost is at most that for the CbCI algorithm given the partitioning of the network, i.e., $O(N^2\log N)$. Therefore, if the CbCI algorithm is naively implemented, community detection is not a bottleneck in terms of the computation time when any of these five community detection algorithms is used.
If $N_{\rm C}=O(N)$ and we implement the CbCI algorithm using the max-heap data structure, a community detection algorithm requiring more than $O(N\log N)$ time presents a bottleneck. In this case, the Infomap when the network is sparse (i.e., $M=O(N)$), label-propagation algorithm, and Louvain algorithm retain $O(N\log N)$ total computation time of the CbCI algorithm. The total computation time with any of the other three community detection algorithms is governed by that of the community detection algorithm.
      
\section*{Results}
 
In this section, we compare the performance of the CbCI algorithm with the CI and other immunization algorithms (see Methods) on two model networks and 12 empirical networks.
Let $q$ be the fraction of removed nodes.
The size of the LCC after $qN$ nodes have been removed, divided by $N$, is denoted by $G(q)$. 

\subsection*{Scale-free network models with and without community structure}
  
\begin{figure}
\includegraphics[width=1\columnwidth,clip]{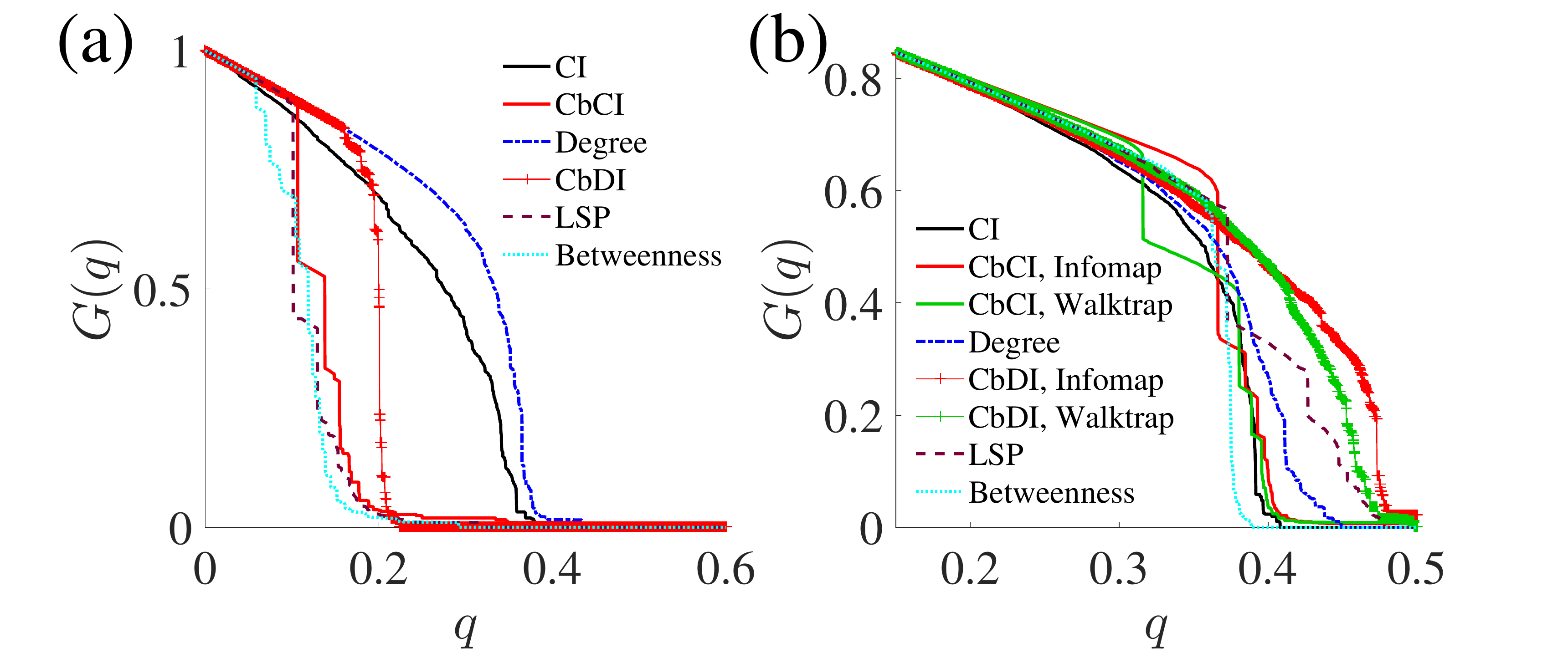}
\caption{Normalized size of the LCC, $G(q)$, plotted against the fraction of removed nodes, $q$, in model networks with $N=5000$. A curve corresponds to an immunization algorithm. See Methods for the abbreviations. 
(a) Scale-free network with prescribed community structure. (b) BA model.}
\label{fig:BAmodel}
\end{figure}

We start by testing various immunization algorithms on a scale-free network model with built-in community structure (Methods).
We sequentially remove nodes from this network according to each immunization algorithm and track the size of the LCC. We use the community structure imposed by the model to inform the CbCI and CbDI algorithms. The results for a range of immunization algorithms are shown in Fig.~\ref{fig:BAmodel}(a).
Both CbCI and CbDI algorithms considerably outperform the CI algorithm. The CbCI algorithm performs better than the CbDI algorithm. The performance of the CbCI algorithm is close to the Betweenness algorithm. It should be noted that the Betweenness algorithm, while efficient, is not scalable to larger networks.

Next, we consider a scale-free network without community structure, which is generated by the original BA model with $N=5000$ and $\langle k \rangle \approx 12$ (the parameters of the model are equal to $m_0=m=6$). We run the CbCI and CbDI strategies by applying a community detection algorithm to the generated network although the BA model lacks community structure. In fact, all but the label-propagation algorithm returns a partitioning result. The performance of the different immunization algorithms for this network is compared in Fig.~\ref{fig:BAmodel}(b). 
The CbCI algorithm combined with Infomap or Walktrap outperforms the Degree and LSP algorithms. The performance of the CbCI algorithm is close to that of the CI algorithm except in an early stage of node removal. A different community-based immunization algorithm, CbDI, lacks this feature. 
This result suggests that the CbCI algorithm combined with Infomap or Walktrap can work efficiently even when the network does not have community structure.

The results for the CbCI and CbDI algorithms combined with the other four community detection algorithms are shown in Fig.~\ref{fig:comp13data}(a). The figure suggests that the CbCI algorithm combined with Infomap or Walktrap performs better than when it is combined with a different community detection algorithm.
    
\subsection*{Empirical networks}\label{sec:realworld_networks}
 
In this section, we run the CbCI and other algorithms on the following 12 empirical networks with community structure. (i) Two networks of Autonomous Systems of the Internet constructed by the University of Oregon Route Views project~\cite{Leskovec2005HepAS,SNAP,OregonRouteView}: A node is an Autonomous System. The network collected on 2 January 2000 and that on 31 March 2001 are referred to as AS-1 and AS-2, respectively.
%
%
(ii) Pretty Good Privacy network (PGP)~\cite{Boguna2004}: Two persons are connected by a link if they share confidential information using the PGP encryption algorithm on the Internet.
(iii) World Wide Web (WWW)~\cite{Albert1999Nature}: A network of websites connected by hyperlinks, which is originally a directed network. (iv)
Email-based communication network at Kiel University (referred to as email-uni)~\cite{Ebel2002PRE-email}: E-mail sending activity among students, which provides a directed link, recorded over a period of 112 days. (v) Email-based communication network in Enron Corporation (email-Enron)~\cite{Klimt2004enron,Leskovec2009enron,SNAP}: Two e-mail users in the data set are connected by an unweighted directed link if at least one e-mail has been sent from one user to the other user. (vi) Collaboration networks in General Relativity and Quantum Cosmology (CA-GrQc), Astro Physics, (CA-Astroph), and Condensed Matter, (CA-Condmat) categories~\cite{Leskovec2007CA,SNAP} and High Energy Physics -- Phenomenology (CA-HepPh) and
High Energy Physics  -- Theory (CA-HepTh) categories in arXiv~\cite{Leskovec2005HepAS,SNAP}.
By definition, two authors are adjacent if they coauthor a paper.
(vii) High-energy physics citation network within the hep-th category of arXiv (HEP)~\cite{usair-pajek}, which is originally a directed network.
For each network, we removed the link weight, self-loops, and direction of the link, and submitted the LCC to the following analysis.
Summary statistics of these networks including the modularity, $Q$, are shown in Tables~\ref{tab:DataDescription} and~\ref{tab:Nc_Q}.

We do not investigate the Betweenness immunization algorithm due to its high computational cost (i.e., $O(NM)$ time for calculating the betweenness centrality of all nodes \cite{Brandes2001JMathSociol}, hence $O(N^2 M)$ time for removing $O(N)$ nodes).
   
\begin{figure}
\includegraphics[width=1\columnwidth,clip]{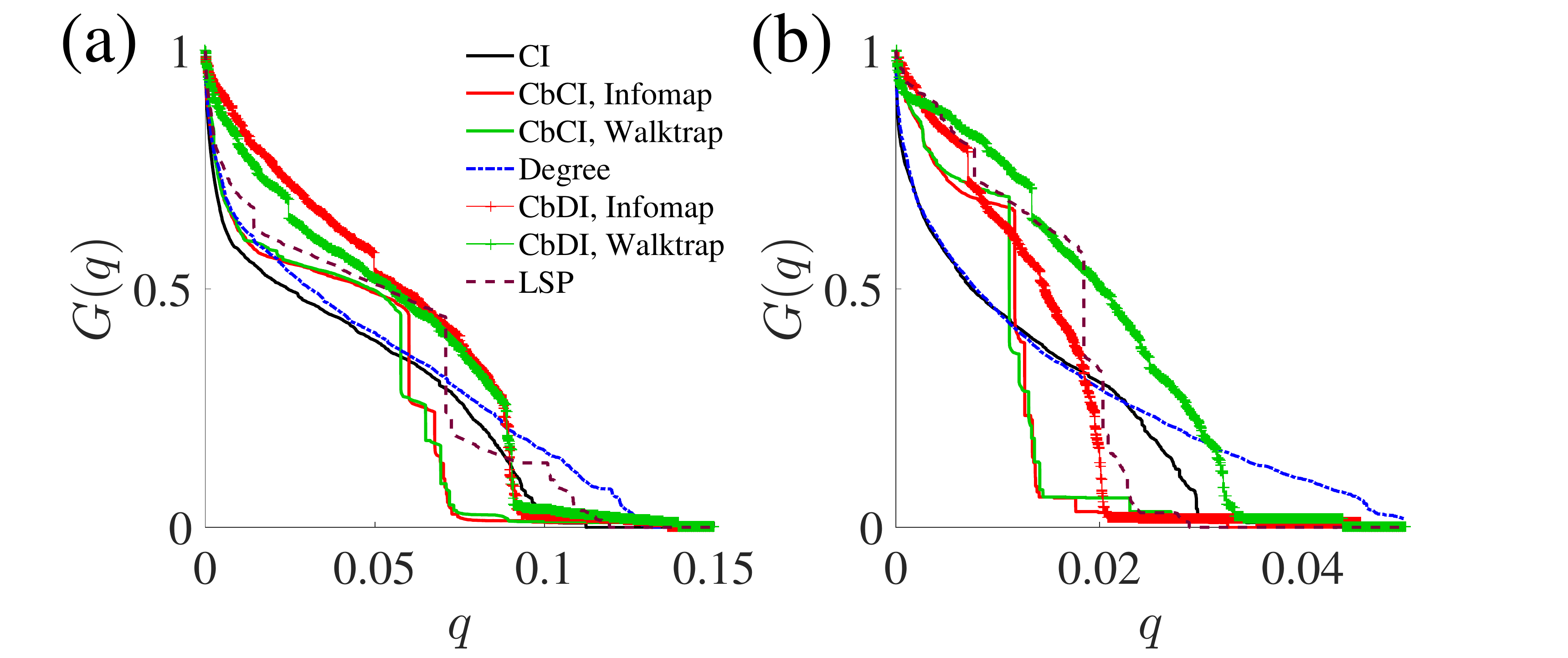}
\caption{Normalized size of the LCC, $G(q)$, plotted against the fraction of removed nodes, $q$, in two empirical networks. (a) E-mail communication network in Enron. (b) World Wide Web.} 
\label{fig:comp_enron_www}
\end{figure}

The performance of the different immunization algorithms is compared on two empirical networks in Fig.~\ref{fig:comp_enron_www}. Among the 12 empirical networks that we tested, these two networks yielded the smallest and largest modularity values as maximized by the Louvain algorithm. The figure indicates that the CbCI algorithm combined with Infomap or Walktrap performs better than the previously proposed algorithms including the CI algorithm in both networks. The CbCI algorithm performs better than the CI algorithm in many other empirical networks as well (Fig.~\ref{fig:comp13data}(b)--(m)). Furthermore, the CbCI algorithm combined with a different community detection algorithm also outperforms the CI algorithm in most of the networks (Fig.~\ref{fig:comp13data}(b)--(m)).
 
To be quantitative, we measure the fraction of removed nodes at which the network fragments into sufficiently small connected components, i.e.,
\begin{equation}
q_{\rm c}\equiv\inf\{q: G(q)<\theta\},
\label{eq:def q_c}
\end{equation}
where we remind that $G(q)$ is the size of the LCC normalized by $N$. We set $\theta=0.05$. We calculate $q_{\rm c}$ for each combination of a network and an immunization algorithm.

\begin{figure}
\includegraphics[width=1\columnwidth,clip]{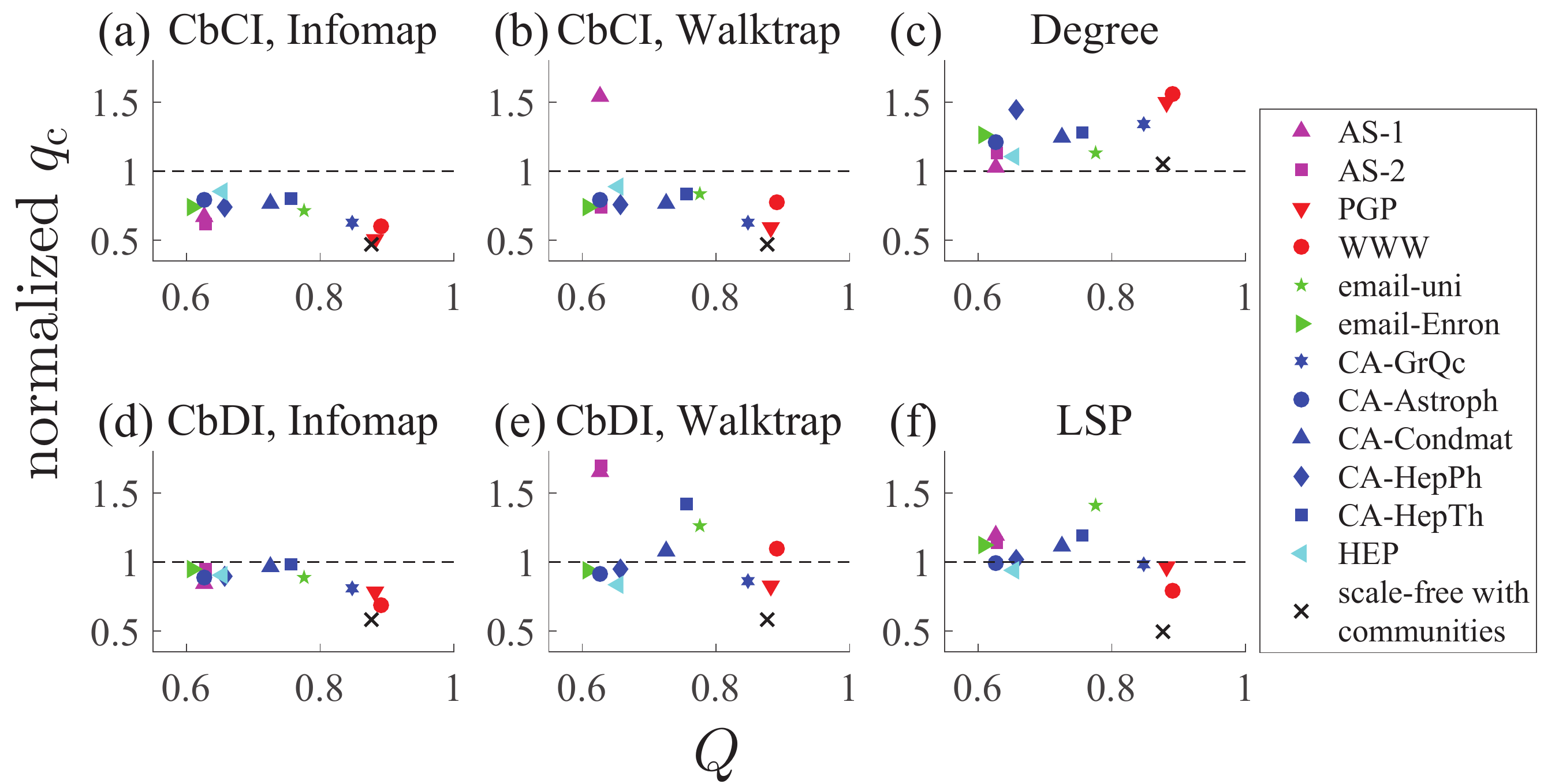}
\caption{The fraction of removed nodes to fragment the network, $q_{\rm c}$, for an immunization algorithm divided by the value for the CI algorithm. (a) CbCI combined with Infomap. (b) CbCI combined with Walktrap. (c) High degree adaptive (Degree). (d) CbDI combined with Infomap. (e) CbDI combined with Walktrap. (f) Laplacian spectral partitioning (LSP). A symbol represents a network. The cross represents the model network used in Fig.~\ref{fig:BAmodel}(a). The modularity value, $Q$, is determined by the Louvain algorithm.}
\label{fig:scatter_main_qc}
\end{figure}

The value of $q_{\rm c}$ for each immunization algorithm normalized by the $q_{\rm c}$ value for the CI algorithm is plotted in Fig.~\ref{fig:scatter_main_qc}. A symbol represents a network. A small normalized value of $q_{\rm c}$ implies a high efficiency of the immunization algorithm. As expected, the Degree immunization algorithm performs worse than the CI in all the tested networks (Fig.~\ref{fig:scatter_main_qc}(c)). 
For the CbCI algorithm combined with Infomap, $q_{\rm c}$ is smaller by 15.0\% to 49.7\% than that for the CI algorithm (Fig.~\ref{fig:scatter_main_qc}(a)). The CbCI algorithm combined with Walktrap shows a similar performance for all but one networks (Fig.~\ref{fig:scatter_main_qc}(b)). The CbCI algorithm combined with three of the other four community detection algorithms performs better than the CI algorithm for networks with relatively strong community structure
(Fig.~\ref{fig:scatter_SI_qc}). The CbDI algorithm combined with Infomap performs better than the CI algorithm for all networks, but to a lesser extent than the CbCI algorithm combined with Infomap does (Fig.~\ref{fig:scatter_main_qc}(d)). The CbDI algorithm combined with Walktrap (Fig.~\ref{fig:scatter_main_qc}(e)) and the other four community detection algorithms (Fig.~\ref{fig:scatter_SI_qc}) performs worse than the CI algorithm. The LSP algorithm performs worse than the CI algorithm in a majority of the networks (Fig.~\ref{fig:scatter_main_qc}(f)).

Even if two immunization algorithms yield the same $q_{\rm c}$ value on the same network, $G(q)$ may considerably drop at a smaller $q$ value with one immunization algorithm than the other algorithm. To quantify the performance of immunization algorithms in this sense, we measure the size of the LCC integrated over $q$ values~\cite{Schneider2011PRE,Schneider2011PNAS}, i.e.,
  \begin{align}\label{eq:def_measure_ave}
      \overline{G} \equiv \frac{1}{N}\sum_{i=0}^N G(i/N).
  \end{align}
It should be noted that $\overline{G}$ is the area under the curve when $G(q)$ is plotted against $q$ and ranges between 0 and $1/2$.
A small $\overline{G}$ value implies a good performance of an immunization algorithm. 

\begin{figure}
\includegraphics[width=1.0\columnwidth,clip]{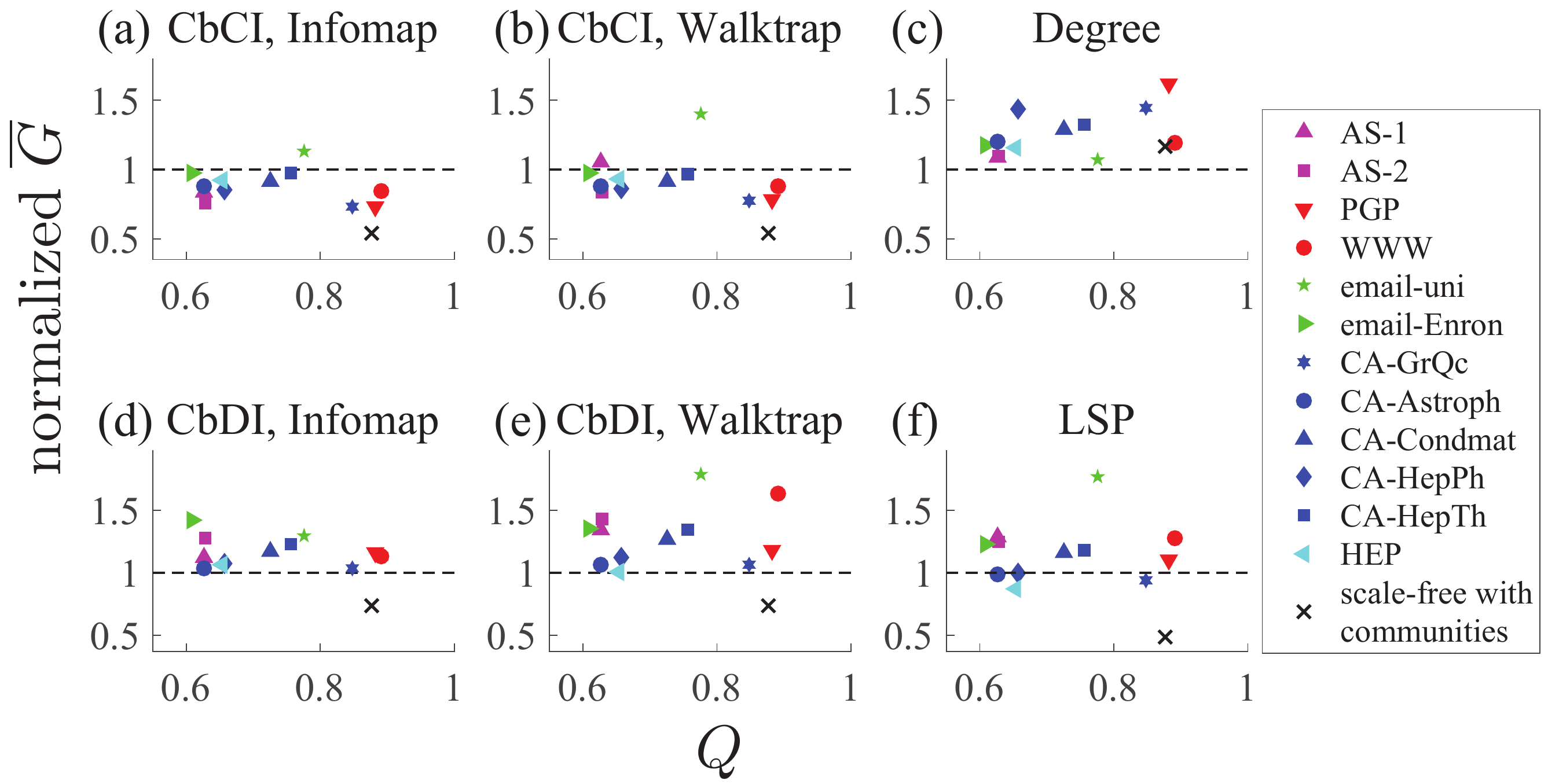}
\caption{The $\overline{G}$ value normalized by that for the CI algorithm. (a) CbCI combined with Infomap. (b) CbCI combined with Walktrap. (c) Degree. (d) CbDI combined with Infomap. (e) CbDI combined with Walktrap. (f) LSP. The $Q$ value is determined by the Louvain algorithm.}
\label{fig:scatter_main_G}
\end{figure}

The value of $\overline{G}$ for each immunization algorithm normalized by that for the CI algorithm is plotted in 
Fig.~\ref{fig:scatter_main_G}. The CbCI algorithm combined with Infomap outperforms the CI algorithm in 11 out of the 12 networks in terms of $\overline{G}$ (Fig.~\ref{fig:scatter_main_G}(a)). Similarly, the CbCI algorithm combined with Walktrap outperforms the CI algorithm in ten out of the 12 networks (Fig.~\ref{fig:scatter_main_G}(b)). The CbCI combined with any of the other four community detection algorithms outperforms the CI algorithm in roughly half of the networks and tends to be efficient for networks having large modularity values as determined by the Louvain algorithm (Fig.~\ref{fig:scatter_SI_G}).
In particular, for the three networks with the largest modularity, the CbCI algorithm combined with any of the six community detection algorithms outperforms the CI algorithm. The Degree, CbDI, and LSP algorithms are less efficient than the CI algorithm in terms of $\overline{G}$ (Figs.~\ref{fig:scatter_main_G}(c)--(f) and~\ref{fig:scatter_SI_G}).

\subsection*{Why do Infomap and Walktrap marry better with the CbCI algorithm than the other community detection algorithms?}

We have shown that the CbCI algorithm is more efficient when it is combined with
Infomap or Walktrap, in particular Infomap, than with the other four community detection algorithms. To explore why, we start by measuring the clustering coefficient \cite{Watts1998Nature} of the unweighted version of the coarse-grained networks. We do so because in theory the CI assumes locally tree-like networks \cite{Morone2015Nature,Radicchi2016PRER}. High clustering in the coarse-grained network may discourage the CbCI algorithm. For each empirical network, we measure the Pearson correlation coefficient between the clustering coefficient and $q_{\rm c}$ normalized by the value for the CI algorithm. We use the result for each community detection algorithm as a data point such that the correlation coefficient is calculated on the basis of six data points. The results are shown in Table~\ref{tab:corr_givennet_qc}.
%
%
We find that the clustering coefficient is not consistently correlated with 
the normalized $q_{\rm c}$. The results are qualitatively the same with a weighted clustering coefficient~\cite{OnnelaPRE2005,Saramaki2007PRE} (Table~\ref{tab:corr_givennet_qc}). We obtain similar results if 
$\overline{G}$ instead of $q_{\rm c}$ is used as a performance measure (Table~\ref{tab:corr_givennet_G}). It should be noted that different community detection algorithms yield sufficiently different clustering coefficient values including large values (Fig.~\ref{fig:regress_givennet}(a)). We conclude that the lack of local tree-like structure in the coarse-grained networks is not a strong determinant of the performance of the CbCI algorithm. This result does not contradict those for the original CI algorithm, which assumes local tree-like networks, because the CI algorithm is practically efficient on loopy networks as well \cite{Morone2015Nature}.

We have set $\ell=2$, thus ignoring the contribution of nodes in coarse-grained networks three or more hops away from a focal node. In fact, large coarse-grained networks may have a large mean path length and deteriorate the performance of the CbCI algorithm. Therefore, we calculate the correlation coefficient between $N_{\rm C}$, i.e., the number of the detected communities, and $q_{\rm C}$, and between the mean path length in the unweighted coarse-grained network and $q_{\rm C}$ (Table~\ref{tab:corr_givennet_qc}). The correlation efficient between $\overline{G}$ and either $N_{\rm C}$ or the mean path length is also measured (Table~\ref{tab:corr_givennet_G}).
The tables indicate that the performance of a community detection algorithm is not consistently correlated with the mean path length.
It is correlated with $N_{\rm C}$, but in the manner such that the performance of the CbCI algorithm improves as $N_{\rm C}$ increases, contrary to the aforementioned postulated mechanism. Therefore, the use of $\ell=2$ does not probably explain the reason why a community detection algorithm marries the CbCI algorithm better than another.

In fact, the CbCI algorithm performs well when the detected communities have relatively similar sizes. To show this, we measure the entropy in the partitioning, which is defined by $\sum_{c=1}^{N_{\rm C}}(N^{\prime}_c/N) \log (N^{\prime}_c/N)$, where $N^{\prime}_c$ is the number of nodes in the $c$th community. The entropy ranges between 0 and $\log N_{\rm C}$. A large entropy value implies that the partitioning of the network is relatively egalitarian. The correlation coefficient between the entropy and the normalized $q_{\rm c}$ is shown in 
Table~\ref{tab:corr_givennet_qc} for each network. The entropy and $q_{\rm c}$ are  negatively correlated with each other for all networks and strongly so for most of the networks. This result is robust when we normalize the entropy by the largest possible value, i.e., $\log N_{\rm C}$ (Table~\ref{tab:corr_givennet_qc}), and when the performance measure is replaced by $\overline{G}$ (Table~\ref{tab:corr_givennet_G}). 

To assess the robustness of this finding, we calculate the same correlation coefficient between either the unnormalized or normalized entropy and one of the two performance measures, but for each community detection algorithm. Now each empirical network constitutes a data point based on which the correlation coefficient is calculated. The correlation coefficient values are shown in Table~\ref{tab:corr_givenalg}. Although the correlation is weaker than in the previous case, the correlation between the entropy and either the normalized $q_{\rm C}$ or $\overline{G}$ is largely negative, which is consistent with the results shown in Tables~\ref{tab:corr_givennet_qc} and \ref{tab:corr_givennet_G}. The correlation coefficient between $Q$ and each of the performance measure is also shown in Table~\ref{tab:corr_givenalg}. The entropy provides a weaker determinant of the performance as compared to $Q$, which is expected because the CbCI algorithm is designed for networks with community structure. Nevertheless, the entropy provides a larger (i.e., more negative) correlation value than $Q$ does in some cases (Table~\ref{tab:corr_givenalg}).

Infomap tends to detect a large number of communities 
(Table \ref{tab:Nc_Q}) whose size is less heterogeneously distributed than the case of the other community detection algorithms (Figs.~\ref{fig:regress_givennet}(i) and (k)).
 We consider that this is a main reason why Infomap is effective when combined with the CbCI algorithm. Roughly speaking, the label-propagation algorithm tends to yield a similarly large number of communities, $N_{\rm C}$ (Table \ref{tab:Nc_Q}). However, the size of the community is more heterogeneously distributed with the label-propagation algorithm than with Infomap, as quantified by the entropy measures (Figs.~\ref{fig:regress_givennet}(i) and (k)).
 
\section*{Discussion}

We showed that the CbCI immunization algorithm outperforms the CI and some other algorithms when a given network has community structure. The algorithm aims to pinpoint nodes that connect different communities at a reasonable computational cost.
The CbCI algorithm is in particular efficient when Infomap \cite{Rosvall2008,Rosvall2010} is used for detecting communities beforehand. Infomap runs sufficiently fast at least for sparse networks \cite{Fortunato2010PhysRep} such that the entire CbCI algorithm runs as fast as the CI algorithm at least asymptotically in terms of the network size. The Walktrap community detection algorithm \cite{Pons2005} is the second best among the six candidates to be combined with the CbCI algorithm in terms of the quality of immunization. However, Walktrap is slower than Infomap. Walktrap consumes longer time than the main part of the CbCI algorithm, i.e., sequential node removal, when the max-heap data structure is used for implementing the CbCI algorithm.
In this case, the community detection before starting the node removal is the bottleneck of the entire CbCI algorithm, and the CbCI algorithm is slower than the CI algorithm. To our numerical efforts, we recommend Infomap to be combined with the CbCI algorithm.

We argued that Infomap works better in combination with the CbCI algorithm than the other community detection algorithms do mainly because Infomap yields a relatively egalitarian distribution of the community size. However, the distribution of the community size is usually skewed even with Infomap~\cite{Lancichinetti2010Plosone}. The CbCI algorithm may work even better if we use a community detection algorithm that imposes that the detected communities are of the equal or similar sizes. This problem is known as $k$-balanced partitioning, where $k$ refers to the number of communities. Although $k$-balanced partitioning for general $k$ is notoriously hard to solve, there are various approximate algorithms for this problem \cite{Andreev2006BGP,Krauthgamer2009SODA,Feldmann2015Algo}.
Combining these algorithms with the CbCI algorithm may be profitable.

We partitioned the network just once in the beginning of the CbCI algorithm and used the obtained community structure throughout the node removal procedure. This property is shared by the CbDI algorithm \cite{Masuda2009NJP} and another immunization algorithm~\cite{Requiao2015Plosone}. We may be able to improve the performance of immunization by updating the community structure during the node removal. Our preliminary numerical simulations did not yield an improvement of the CbCI algorithm with online updating of community structure (section~\ref{sec:community_update} in the SI). We should also bear in mind the computational cost of community detection, which would be repeatedly applied in the case of online updating. Nevertheless, this line of improvement may be worth investigating. 
 
The CI assumes locally tree-like networks \cite{Morone2015Nature}. Although the CI algorithm is practically efficient in moderately loopy networks as well \cite{Morone2015Nature}, many empirical networks are abundant in triangles and short cycles such that they are highly loopy \cite{Newman2010book}. Dense connectivity within a community implies that there tend to be many triangles and short cycles in a network with community structure \cite{Radicchi2004PNAS,Palla2005Nature}. Then, coarse graining effectively coalesces many triangles and short cycles into one supernode, possibly suppressing their detrimental effects. At the same time, however, coarse-grained networks tend to have a large clustering coefficient (Fig.~\ref{fig:regress_givennet}(a)). We may be able to improve the performance of the CbCI algorithm by suppressing the effect of short cycles in coarse-grained networks.
Recently, a method has been proposed to improve the accuracy of estimating the percolation threshold using non-backtracking matrices, where redundant paths are suppressed in the counting of the paths \cite{Radicchi2016PRER}. This method applied to both CI and CbCI algorithms may enhance their performance in the immunization problem.

The recently proposed collective influence propagation (CI$_{\rm p}$) algorithm, which can be interpreted as the CI algorithm in the limit of $\ell\to\infty$, generally yields better solutions than the CI algorithm does \cite{Morone2016SciRep}. Given that we have not implemented the CI$_{\rm p}$ algorithm in the present article, we are not arguing that the CbCI algorithm is better than the CI$_{\rm p}$ algorithm. It should also be noted that we may be able to combine the CbCI algorithm with the idea of the CI$_{\rm p}$ algorithm (i.e., using the leading left and right eigenvectors of the non-backtracking matrix) to devise a new algorithm.

\section*{Methods}

\subsection*{Immunization algorithms to be compared\label{sub:strategies tried}}
 
We compare the performance of the CI and CbCI algorithms against the following immunization algorithms.

\begin{itemize}

\item High degree adaptive (abbreviated as Degree)~\cite{Albert2000Nature,Callaway2000PhysRevLett,Cohen2001PhysRevLett}: We sequentially remove the node having the largest degree. If multiple nodes have the largest degree, we break the tie by selecting one of the largest-degree nodes at random. We recalculate the degree after each node has been removed.
   
\item Community-based dynamical importance (CbDI)~\cite{Masuda2009NJP}: This method exploits the community structure of a network, similar to the CbCI algorithm, but calculates the importance of a community in the coarse-grained network in terms of the so-called dynamical importance~\cite{Restrepo2008}. The CbDI algorithm needs a community detection algorithm. We use each of the six community detection algorithms used in the CbCI algorithm. 
%

The CbDI algorithm runs as follows~\cite{Masuda2009NJP}. We denote by $\tilde{\lambda}$ and $\tilde{\bm u} = (\tilde{u}_1\; \cdots\; \tilde{u}_{N_{\rm C}})^{\top}$ the largest eigenvalue and the corresponding eigenvector of $\tilde{A}$, respectively. Owing to the Perron-Frobenius theorem, it holds true that $\tilde{\lambda}>0$ and $\tilde{u}_i \ge 0$ ($1\le i\le N_{\rm C}$). The number of links between node $i$ and to the $J$th community is denoted by $k_{iJ} \equiv \sum_{j\in \text{ community } J} A_{ij}$. We define $x= \left(\sum_{J=1, J\neq I}^{N_{\rm C}}k_{iJ}\tilde{u}_J\right)/\tilde{\lambda}$, where $I$ is the community to which node $i$ belongs. The CbDI of node $i$ is defined by $(2\tilde{u}_I - x)\sum_{J=1, J\neq I}^{N_{\rm C}} k_{iJ} \tilde{u}_J$. We remove the nodes in descending order of the CbDI. If there are multiple nodes that have the same largest CbDI value, we break the tie by selecting the node that has the largest number of intra-community links. We recalculate the CbDI values of all the remaining nodes after removing each node. Once all the communities are disconnected, we sequentially remove the nodes in descending order of $k_{iI}$. We recalculate $k_{iI}$ of all the remaining nodes after removing each node.

\item The Laplacian spectral partitioning (LSP) algorithm runs as follows~\cite{Zahedi2015IKT}:

\begin{enumerate}

    \item For the largest connected component (LCC), calculate the Fiedler vector,  i.e., the eigenvector associated with the smallest positive eigenvalue of the Laplacian, $L \equiv D_{\rm LCC} - A_{\rm LCC}$, where $D_{\rm LCC}$ denotes the $N_{\rm LCC}\times N_{\rm LCC}$ diagonal matrix whose $(i,i)$ element is equal to the degree of the $i$th node in the LCC, $N_{\rm LCC}$ is the number of nodes in the LCC, and $A_{\rm LCC}$ is the adjacency matrix of the LCC.

    \item Partition the $N_{\rm LCC}$ nodes into two non-empty groups by thresholding on the value of the element in the Fiedler vector. Group 1 (group 2) consists of the nodes whose corresponding element in the Fiedler vector is higher (lower) than a threshold. There are $N_{\rm LCC}-1$ possible ways to bipartition the nodes.
    
    \item Calculate 
    \begin{align}
     \mathcal{Q} = m_{\rm{in}}\ln{\left( \frac{2m_{\rm{in}}}{K_1^2+K_2^2}\right)} + m_{\rm out}\ln{\left( \frac{m_{\rm{out}}}{K_1 K_2}\right) },
    \end{align}
for each bipartition, where $m_{\rm{in}}$ and $m_{\rm{out}}$ are the numbers of intra-group and inter-group links, respectively. $K_1$ and $K_2$ represent the sum of the nodes' degrees in groups 1 and 2, respectively.
    
    \item Find the partition that maximizes $\mathcal{Q}$.
    
    \item Given the partition, remove the node that has the largest number of inter-group links. Then, recalculate the number of inter-group links for each remaining node. Repeat the node removal until the two groups are disconnected.  
    
    \item Repeat steps 1--5 until the size of the LCC becomes less than $\theta N$, where $\theta = 0.01$.
        
\end{enumerate}

\item High betweenness centrality adaptive (abbreviated as Betweenness) \cite{Holme2002PRE_attack,Ueno2008JTB,Salathe2010PlosCB,Hebert2013Scirep}: We remove the node with the largest betweenness centrality. If multiple nodes have the same largest betweenness centrality value, the node having the largest degree is removed. We recalculate the betweenness of all nodes every time we remove a node.

\end{itemize}

We excluded the dynamical importance \cite{Restrepo2008} because it is less successful than the CI on various networks \cite{Morone2015Nature} and than the CbDI on networks with community structure \cite{Masuda2009NJP}. We also excluded the immunization algorithms on the basis of the PageRank, closeness centrality, and {\it k}-core, which had been shown to be outperformed by the CI algorithm \cite{Morone2015Nature}. This is because these algorithms do not particularly exploit community structure of the network such that there is no reason for believing that they would perform competitively on networks with community structure.

\subsection*{A scale-free network model with community structure}

We constructed a scale-free network with built-in community structure as follows~\cite{Masuda2009NJP}. We first generate a coarse-grained network whose node is regarded as community, using the Barab\'{a}si-Albert (BA) model~\cite{Barabasi1999Science} having $N_{\rm C}=100$ nodes and mean degree six. The initial network is the clique composed of $m_0=3$ nodes, and each added node has $m=3$ links. After generating a coarse-grained network, we assign 50 nodes to each community, resulting in $N=50 \times N_{\rm C}=5000$ nodes in total. For each community, the intra-community network is given by the BA model with $m_0=m=4$, which yields the mean within-community degree equal to
$\langle k\rangle_{\ell} = 2\left[(N-m_0)m + m_0(m_0-1)/2\right]/N = 7.6$.
Additionally, if communities $I$ and $J$ are adjacent in the coarse-grained network, then nodes $i\in I$ and $j\in J$ are connected with probability $\langle k\rangle_g /(6N/N_{\rm{C}})$. This guarantees that a node is adjacent to $\langle k\rangle_g$ nodes in different communities on average. We set $\langle k\rangle_g = 1$.
The mean degree of the entire network is equal to $\langle k\rangle = 8.58 \approx \langle k \rangle_{\ell} + \langle k \rangle_g$.



\section*{Acknowledgements}

N.M. acknowledges the support provided through JST, CREST, and JST, ERATO, Kawarabayashi Large Graph Project. 
T.K. acknowledges financial support from the Japan Society for the Promotion of Science KAKENHI Grants no.~25780203, 15H01948, and 16K03551. 
We thank Flaviano Morone and Taro Takaguchi for providing codes for the CI algorithm. 

\section*{Author contributions}

T.K. and N.M. conceived the research. T.K. conducted the analysis. T.K. and N.M. discussed the results and wrote the manuscript. 

\section*{Competing financial interests statement}

The authors declare no competing financial interests.

\clearpage

\clearpage

\clearpage

\clearpage

\clearpage

\clearpage
 \begin{table}
  \caption{Correlation coefficient between an explanatory variable and the normalized $q_{\rm c}$. The clustering coefficient is defined by the number of triangles containing the $i$th node divided by $k_i(k_i-1)/2$, which is averaged over all nodes $1\le i\le N$. The weighted clustering coefficient is defined by $\sum_{j,k=1; j, k\neq i}^N (\hat{w}_{ij}\hat{w}_{jk}\hat{w}_{ki})^{1/3}/[k_{i}(k_{i}-1)]$, which is averaged over $i$~\cite{OnnelaPRE2005,Saramaki2007PRE}.
Here, $\hat{w}_{ij} = w_{ij}/\max_{1\le i^{\prime}, j^{\prime}\le N}(w_{i^{\prime}j^{\prime}})$, and $w_{ij}$ is the weight of the link between the $i$th and $j$th nodes.
%
%
We calculated the correlation coefficient for each network on the basis of the data points obtained from the six community detection algorithms. The scattergrams based on which the correlation coefficient has been calculated are shown in Figs.~\ref{fig:regress_givennet}(a), (c), (e), (g), (i), and (k).}
    \label{tab:corr_givennet_qc}
    \begin{tabular}{lP{2cm}P{2cm}P{2cm}P{2cm}P{2cm}P{2cm}}
    \hline 
    \hline 
    \parbox[c][1.6cm][c]{0cm} 
    & \raisebox{-.2cm}{\shortstack{clustering \\{coefficient}}} 
    & \raisebox{-.5cm}{\shortstack{weighted \\{clustering}\\{coefficient}}} 
    & \raisebox{-.0cm}{$N_{\rm C}$\;\;\:} & \raisebox{-.2cm}{\shortstack{mean path\\{length}}} & \raisebox{-.0cm}{entropy} & \raisebox{-.2cm}{\shortstack{normalized\\{entropy}}}  
     \\\hline
AS-1        & $0.238$  & $0.120$  & $-0.415$ & $-0.354$ & $-0.523$ & $-0.608$     \\
AS-2        & $-0.258$ & $0.119$  & $-0.417$ & $0.065$  & $-0.319$ & $-0.203$ \\
PGP         & $0.298$  & $0.490$  & $-0.603$ & $-0.534$ & $-0.667$ & $-0.781$   \\
WWW         & $-0.005$ & $-0.430$  & $0.306$  & $-0.375$ & $0.216$  & $-0.169$ \\
email-uni   & $0.213$  & $-0.053$  & $-0.362$ & $0.125$  & $-0.446$ & $-0.568$  \\
email-Enron & $-0.278$ & $-0.398$ & $-0.073$ & $-0.136$ & $-0.650$ & $-0.817$ \\
CA-GrQc     & $0.438$  & $0.345$  & $-0.773$ & $-0.458$ & $-0.891$ & $-0.934$    \\
CA-Astroph  & $-0.154$ & $-0.005$ & $-0.406$ & $-0.144$ & $-0.764$ & $-0.826$ \\
CA-Condmat  & $0.121$  & $-0.181$  & $-0.653$ & $0.219$  & $-0.820$ & $-0.918$ \\
CA-HepPh    & $0.729$  & $0.792$  & $-0.845$ & $-0.569$ & $-0.932$ & $-0.781$      \\
CA-HepTh    & $-0.200$ & $-0.118$ & $0.178$  & $0.325$  & $-0.067$ & $-0.320$    \\
HEP         & $0.314$  & $0.204$  & $-0.718$ & $-0.042$ & $-0.842$ & $-0.759$ \\
    \hline 
    \end{tabular}
\end{table}

\clearpage
\begin{table}
    \caption{Correlation coefficient between an explanatory variable and the normalized $\overline{G}$ for each network. We calculate the correlation coefficient for each network on the basis of the data points obtained from the six community detection algorithms. The scattergrams based on which the correlation coefficient has been calculated are shown in Figs.~\ref{fig:regress_givennet}(b), (d), (f), (h), (j), and (l).}
    \label{tab:corr_givennet_G}
    \begin{tabular}{lP{2cm}P{2cm}P{2cm}P{2cm}P{2cm}P{2cm}}
    \hline 
    \hline 
   \parbox[c][1.6cm][c]{0cm} 
    & \raisebox{-.2cm}{\shortstack{clustering \\{coefficient}}} 
    & \raisebox{-.5cm}{\shortstack{weighted \\{clustering}\\{coefficient}}} 
    & \raisebox{-.0cm}{$N_{\rm C}$\;\;\:} & \raisebox{-.2cm}{\shortstack{mean path\\{length}}} & \raisebox{-.0cm}{entropy} & \raisebox{-.2cm}{\shortstack{normalized\\{entropy}}}  
     \\\hline
AS-1        & $0.064$  & $0.284$  & $-0.813 $ & $-0.259 $ & $-0.813 $ & $-0.604 $     \\
AS-2        & $-0.369$ & $0.416$  & $-0.693 $ & $0.105 $  & $-0.660 $ & $-0.449 $ \\
PGP         & $0.084$  & $0.295$  & $ -0.489$ & $-0.586 $ & $-0.603 $ & $-0.766 $   \\
WWW         & $-0.367$ & $0.111$  & $-0.713 $ & $0.281 $  & $-0.820 $ & $-0.694 $ \\
email-uni   & $0.394$  & $0.362$  & $-0.810 $ & $0.043 $  & $-0.823 $ & $-0.595 $  \\
email-Enron & $-0.381$ & $-0.224$  & $-0.474 $ & $-0.258 $ & $-0.804 $ & $-0.798 $  \\
CA-GrQc     & $0.034$  & $0.045$  & $-0.467 $ & $-0.043 $ & $-0.679 $ & $-0.909 $    \\
CA-Astroph  & $-0.313$ & $-0.112$ & $-0.137 $ & $-0.232 $ & $-0.526 $ & $-0.700 $ \\
CA-Condmat  & $0.334$  & $-0.024$  & $-0.706 $ & $0.387 $  & $-0.851 $ & $-0.839 $ \\
CA-HepPh    & $0.227$  & $0.419$  & $-0.424 $ & $-0.214 $ & $-0.607 $ & $-0.685 $      \\
CA-HepTh    & $0.248$  & $0.629$  & $-0.632 $ & $-0.129 $ & $-0.754 $ & $-0.730 $    \\
HEP         & $-0.067$ & $0.400$  & $-0.636 $ & $0.395 $  & $-0.759 $ & $-0.722 $ \\
    \hline 
    \end{tabular}
\end{table}

\clearpage
\begin{table}
 \caption{Correlation coefficient between an explanatory variable and a performance measure for each community detection algorithm. We calculate the correlation coefficient for each community detection algorithm on the basis of the data points obtained from the 12 empirical networks. $q_{\rm c}$ and $\overline{G}$ indicate the values normalized by those for the CI algorithm. The scattergrams based on which the correlation coefficient has been calculated are shown in Fig.~\ref{fig:regress_givenalg_entropy}.}
    \label{tab:corr_givenalg}
    \begin{tabular}{lP{1.8cm}P{1.8cm}P{1.8cm}P{1.8cm}P{1.8cm}P{1.8cm}}
    \hline 
    \hline 
    &\multicolumn{2}{c}{entropy} 
    &\multicolumn{2}{c}{normalized entropy}
    &\multicolumn{2}{c}{$Q$}\\ \cmidrule(lr){2-3}  \cmidrule(lr){4-5} \cmidrule(lr){6-7}  
    & $q_{\rm c}\;\;\;$   
    & $\overline{G}\;\;\;$
    & $q_{\rm c}\;\;\;$   
    & $\overline{G}\;\;\;$
    & $q_{\rm c}\;\;\;$ 
    & $\overline{G}\;\;\;$ \\  \hline
     Infomap     & $-0.088$ & $0.358$  &$-0.017 $ &$-0.206 $ & $-0.576$ & $-0.210$     \\
     Walktrap    & $-0.506$ &  $0.014$ &$-0.434 $ &$-0.201 $ & $-0.348$ & $-0.128$       \\
     label propagation          & $-0.690$ & $-0.659$ &$-0.630 $ &$-0.687 $ & $-0.689$ & $-0.797$       \\
     fast greedy & $-0.211$ & $-0.099$ &$0.067 $  &$-0.081 $ & $-0.015$ & $-0.310$      \\
     simulated annealing          & $-0.695$ &  $0.057$ &$-0.288 $ &$-0.005 $ & $-0.330$ & $-0.391$       \\
     Louvain     & $-0.791$ & $-0.001$ &$-0.267 $ & $-0.370$ & $-0.707$ & $-0.370$     \\
    \hline 
    \end{tabular}
\end{table}

\fontsize{0.01pt}{0.01mm}\selectfont
\figcountSI[\label{fig:CompReptDetect}]
\figcountSI[\label{fig:comp13data}]
\figcountSI[\label{fig:scatter_SI_qc}]
\figcountSI[\label{fig:scatter_SI_G}]
\figcountSI[\label{fig:regress_givennet}]
\figcountSI[\label{fig:regress_givenalg_entropy}]
\tabcountSI[\label{tab:DataDescription}]
\tabcountSI[\label{tab:Nc_Q}]
\seccountSI[\label{sec:community_update}]

\clearpage
\pagenumbering{gobble}
     \newcounter{numpage}
    \newcounter{firstpage}
    \newcounter{lastpage}
    \setcounter{firstpage}{1}
    \setcounter{lastpage}{28}
    \setcounter{numpage}{\value{firstpage}}
    \stepcounter{lastpage}

    \makeatletter
    \@whilenum{\value{numpage}<\value{lastpage}}\do{%
      \includegraphics[page=\value{numpage},bb=70 0 780 790,width=1.47\columnwidth,clip]{srep37778_s1.pdf}\par
      \stepcounter{numpage}}
    \makeatother
    
\end{document}